\documentstyle[12pt]{article}
\begin{document}
\begin{center}
{\bf \LARGE  Poincare algebra in chiral $QCD_2$ }

\vspace{2 cm}

{\bf \Large  Fuad M. Saradzhev}

\vspace{1 cm}

{\small \it  Institute of Physics, Academy of Sciences of
             Azerbaijan, \\
             Huseyn Javid pr. 33, 370143 Baku, Azerbaijan \\}

\end{center}

\vspace{3 cm}

\begin{flushleft}
{\bf Abstract}
\end{flushleft}

\rm
For the chiral $QCD_2$ on a cylinder, we give a construction
of a quantum theory consistent with anomaly. We construct the
algebra of the Poincare generators and show that it differs
from the Poincare one.

\newpage

$1.$ Chiral gauge models with right--left asymmetric coupling
of the matter and gauge fields are anomalous. Some of the first--
class constraints at the classical level become second--class
ones at the quantum level. The anomaly raises some problems.
First of all, the anomalous behaviour of the constraints should
be taken into account when we quantize the anomalous model
and construct its quantum theory. Secondly, there is a problem
of relativistic invariance, namely whether the Poincare algebra
is valid in the quantum theory consistent with the anomaly.

The chiral $QED_2$ (the chiral Schwinger model) ~\cite{jack85}
is the simplest example of the anomalous models. There are different
ways of the consistent canonical quantization of this model
~\cite{fadd86}, ~\cite{hall86}, ~\cite{sara91}. In the physical
sector, the corresponding
quantum theory turns out to be relativistically non--invariant.

In this paper, we consider another anomalous model -- the chiral
$QCD_2$. We assume that space is a circle of length ${\rm L}$ ,
$- \frac{\rm L}{2} \leq x < \frac{\rm L}{2} $, so space--time
manifold is a cylinder $S^1 \times R^1$. Our aim is to construct the
quantum theory of the chiral $QCD_2$ and to derive the algebra
of the Poincare generators.

We use the canonical Hamiltonian formalism. For the standard
non--anomalous $QCD_2$, a construction of the quantum theory was
given in ~\cite{lang94}. In our case, to incorporate the anomaly
into this construction we apply the Gupta-- Bleuler method.

\vspace{1 cm}

$2.$  We consider the most general version of the chiral $QCD_2$,
namely the model in which the right--handed and left--handed
components of the massless Dirac field are coupled to two different
Yang--Mills fields. With $A_{\mu}^{\pm} = A_{\mu}^{\pm, a} \frac{1}{2}
{\tau}^a$ the Yang--Mills fields and ${\psi}_{\pm} = \frac{1}{2}
(1 \pm {\gamma}^5) \psi $ the Dirac fields, the Lagrangian density
for our model is
\[
{\cal L} = {\cal L}_{+} + {\cal L}_{-} ,
\]
\begin{equation}
{\cal L}_{\pm} = - \frac{1}{2} {\rm tr}({\rm F}_{\mu \nu}^{\pm}
{\rm F}_{\pm}^{\mu \nu}) + i {\bar{\psi}}_{\pm} {\gamma}^{\nu}
{\rm D}_{\nu}^{\pm} {\psi}_{\pm} ,  \\
\label{eq: odin}
\end {equation}
where ${\rm D}_{\nu}^{\pm} {\psi}_{\pm} = ({\partial}_{\nu} -
i e_{\pm} A_{\nu,\pm}) {\psi}_{\pm}$ , $\bar{\psi}_{\pm}=
{\psi}_{\pm}^{\dagger} {\gamma}^0$ , ${\tau}_{a}$
$(a={\overline{1,3}})$ are the Pauli matrices, $e_{\pm}$ are the
coupling constants, and ${\rm F}_{\mu,\nu}^{\pm}=
{\partial}_{\mu} A_{\nu}^{\pm} - {\partial}_{\nu} A_{\mu}^{\pm}
- i e_{\pm} [A_{\mu}^{\pm}, A_{\nu}^{\pm}]_{-}$ are the YM field
strength tensors.

We choose the Dirac matrices as ${\gamma}^0 ={\tau}_1$,
${\gamma}^1 = -i {\tau}_{2}$, ${\gamma}^5 = {\gamma}^0
{\gamma}^1 = {\tau}_{3}$. The structure group of the YM fields
is $SU(2)$ and $(\frac{1}{2} {\tau}^a)$ are the generators of the
corresponding Lie algebra in the fundamental representation of
the group.

For $e_{+}=e_{-} \equiv e\sqrt{2}$ and $A_{\mu}^{+}=A_{\mu}^{-}
\equiv \frac{1}{\sqrt{2}} A_{\mu}$, we get from ~\ref{eq: odin}
the Lagrangian density of the standard $QCD_2$ with the Dirac
field ${\psi}$ coupled to the YM field $A_{\mu}$. For $e_{+}=0$,
$A_{\mu}^{+}=0$ ( or $e_{-}=0, A_{\mu}^{-}=0$ ), we have the model
in which the YM field is coupled only to one chiral component of
the Dirac field.

The classical Hamiltonian density is
\[
{\cal H} = {\cal H}_{+} + {\cal H}_{-},
\]
\[
{\cal H}_{\pm} = {\cal H}_{\rm YM}^{\pm} + {\cal H}_{\rm F}^{\pm}
- A_{0,\pm}^{a} {\rm G}_{\pm}^{a},
\]
where ${\cal H}_{\rm YM}^{\pm} = \frac{1}{2} ({\Pi}_{1,\pm}^a)^2$,
with ${\Pi}_{1,\pm}^a$ the momenta canonically conjugate to
$A_{1,\pm}^a$,
\[
{\cal H}_{\rm F}^{\pm} = {\cal H}_{0}^{\pm} \mp e_{\pm} j_{\pm}^a
A_{1,\pm}^a ,
\]
with ${\cal H}_{0}^{\pm} = \mp i {\psi}_{\pm}^{\dagger} {\partial}_1
{\psi}_{\pm}$ the free fermionic Hamiltonian densities,
\[
j_{\pm}^a = {\psi}_{\pm}^{\dagger} \frac{1}{2} {\tau}^a {\psi}_{\pm}
\]
are the fermionic currents, and
\[
{\rm G}_{\pm} = {\rm D}_{1} {\Pi}_{1,\pm} + e_{\pm} j_{\pm}
\]
are the Gauss law generators, $({\rm D}_{1} {\Pi}_{1,\pm})^{a}
\equiv {\partial}_{1} {\Pi}_{1,\pm}^a + e_{\pm} {\varepsilon}_{abc}
A_{1,\pm}^{b} {\Pi}_{1,\pm}^{c}$.

Note that ${\Pi}_{0,\pm}^a =0$ are the primary constraints which
imply the secondary ones ${\rm G}_{\pm}^a=0$. In what follows we
will use the temporal gauge $A_{0,\pm}^a=0$.

Two other generators of the Poincare algebra, i.e. the momentum
and the boost generator, are given by
\begin{eqnarray*}
{\cal P}_{\pm} & = & -i{\psi}_{\pm}^{\dagger} {\partial}_{1}
{\psi}_{\pm} - {\Pi}_{1,a}^{\pm} {\partial}_{1} A_{1,\pm}^a, \\
{\cal K}_{\pm} & = & x {\cal H}_{\pm}. \\
\end{eqnarray*}
On the constrained submanifold ${\rm G}_{\pm}^a =0$, we get
\[
{\cal P}_{\pm} = \pm {\cal H}_{\rm F}^{\pm}.
\]

On the circle boundary conditions for the fields must be specified.
We impose the periodic ones
\begin{eqnarray*}
A_{1,\pm}^a(-\frac{\rm L}{2}) & = & A_{1,\pm}^a(\frac{\rm L}{2}),\\
{\psi}_{\pm}(-\frac{\rm L}{2}) & = & {\psi}_{\pm}(\frac{\rm L}{2}). \\
\end{eqnarray*}
We require also that ${\cal H}_{\rm YM}^{\pm}$ and ${\cal H}_{\rm F}^
{\pm}$ be periodic. Without loss of generality, we can put
\begin{equation}
{\cal H}_{\rm YM}^{\pm}(\frac{\rm L}{2}) =
{\cal H}_{\rm F}^{\pm}(\frac{\rm L}{2}) = 0 . \\
\label{eq: dva}
\end{equation}

Next we transform the fields to their momentum representation which
on the circle is discrete. We get
\begin{eqnarray*}
A_{1,\pm}^a(x) & = & \sum_{n \in \cal Z} A_{1,\pm}^a(n)
e^{i\frac{2\pi}{\rm L} nx} , \\
{\psi}_{\pm}(x) & = & \frac{1}{\sqrt{\rm L}} \sum_{n \in \cal Z}
{\psi}_{\pm}(n) e^{i\frac{2\pi}{\rm L} nx},\\
\end{eqnarray*}
and in all other cases
\[
{\rm X}(x) = \frac{1}{\rm L} \sum_{n \in \cal Z}
{\rm X}(n) e^{i\frac{2\pi}{\rm L} nx}
\]
for ${\rm X}= {\Pi}_{1,\pm} , j_{\pm}, {\cal H}_{0}^{\pm},
{\cal P}^{\pm}, {\cal K}^{\pm}$.

\vspace{1 cm}

$3.$ At the quantum level the fields are represented by operators
which act on a Hilbert space. The canonical commutation relations
for the Fourier transformed field operators are
\[
[\hat{A}_{1,\pm}^a(n) , \hat{\Pi}_{1,\pm}^b(m)]_{-}  =  i
\delta^{ab} \delta_{n,-m},
\]
\begin{equation}
[\hat{\psi}_{\pm}(n) , \hat{\psi}_{\pm}^{\dagger}(m)]_{+}  =
\delta_{n,m}.
\label{eq: tri}
\end{equation}
We assume that the Hilbert space is a fermionic Fock space with
vacuum $|{\rm vac}; A \rangle$ such that
\begin{eqnarray*}
\hat{\psi}_{+}(n) |{\rm vac}; A \rangle =0 & {\rm for} & n>0, \\
\hat{\psi}_{+}^{\dagger}(n) |{\rm vac};A \rangle =0 & {\rm for} & n 
\leq 0,\\
\end{eqnarray*}
and
\begin{eqnarray*}
\hat{\psi}_{-}(n) |{\rm vac}; A \rangle =0 & {\rm for} & n \leq 0, \\
\hat{\psi}_{-}^{\dagger}(n) |{\rm vac};A \rangle =0 & {\rm for} & 
n>0. \\
\end{eqnarray*}
At the same time, the fermionic Fock states are functionals of
$A_{1,\pm}^a(n)$ with $\hat{\Pi}_{1,\pm}^a(n) =
- i {\partial}/{\partial}A_{1,\pm}^a(-n)$ .

The fermionic currents, the Hamiltonian densities and other fermionic
bilinears should be normal ordered $: {\cdots} :$ with respect to
the vacuum $|{\rm vac};A\rangle$. This modifies their naive 
commutation
relations following from ~\ref{eq: tri} as Schwinger terms show up.
In the momentum representation, we have ~\cite{lang94}, 
~\cite{godd86},
~\cite{care87}
\begin{equation}
[\hat{j}_{\pm}^a(n) , \hat{j}_{\pm}^b(m)]_{-} =
i {\varepsilon}_{abc} \hat{j}_{\pm}^c(n+m) \pm n \delta_{n,-m}
\delta_{ab}, \\
\label{eq: cet}
\end{equation}
\begin{equation}
[\hat{\cal H}_{0}^{\pm}(n), \hat{\cal H}_{0}^{\pm}(m)]_{-} =
\pm \frac{2\pi}{\rm L} (n-m) \hat{\cal H}_{0}^{\pm}(n+m)
\pm \frac{1}{3} (\frac{2\pi}{\rm L})^2 n(n^2-1) \delta_{n,-m}, \\
\label{eq: pet}
\end{equation}
where the second term on the r.h.s. of ~\ref{eq: cet} and
~\ref{eq: pet} is the Kac--Moody and Virasoro cocycles, respectively,
and
\[
[\hat{\cal H}_{0}^{\pm}(n) , \hat{j}_{\pm}^a(m)]_{-} =
\mp \frac{2\pi}{\rm L} m \hat{j}_{\pm}^a(n+m),
\]
with no Schwinger term arising here.

The Fourier transformed Gauss law generators are
\[
\hat{\rm G}_{\pm}^a(n) = i \frac{2\pi}{\rm L} n \hat{\Pi}_{1,\pm}^a(n)
+ e_{\pm} {\varepsilon}_{abc} \sum_{p \in \cal Z}
A_{1,\pm}^b(n+p) \hat{\Pi}_{1,\pm}^c(-p) + e_{\pm} 
\hat{j}_{\pm}^a(n). 
\]
The anomaly appears as a central charge for the commutation algebra
of the generators $\hat{\rm G}_{\pm}^a(n)$. Indeed, we have
\begin{equation}
[\hat{\rm G}_{\pm}^a(n) , \hat{\rm G}_{\pm}^b(m)]_{-} =
i e_{\pm} {\varepsilon}_{abc} \hat{\rm G}_{\pm}^c(n+m)
\pm e_{\pm}^2 n \delta_{n,-m} \delta^{ab} ,  \\
\label{eq: shest}
\end{equation}
i.e. the generators $\hat{\rm G}_{+}^a(n)$ and $\hat{\rm G}_{-}^a(n)$
form a Kac--Moody algebra with positive and negative central charge
correspondingly. This central charge destroys the first--class nature
of the constraints and all constraints with non--zero Fourier index
become second--class ones.

For the standard $QCD_2$ with $e_{+}=e_{-}$, the commutation algebra
of the total generators $\hat{\rm G}^a(n) = \hat{\rm G}_{+}^a(n) +
\hat{\rm G}_{-}^a(n)$ has vanishing central charge and so no anomaly.

In terms of states in Hilbert space, the nonvanishing central
charge in ~\ref{eq: shest} means that the local gauge symmetry
is realized projectively ~\cite{jack83} and that we can not 
define physical states as those which are annihilated by the Gauss 
law generators.

For the chiral Schwinger model, the gauge symmetry is abelian
and the Gauss law generators are therefore scalars. This makes
the anomalous behaviour of the model trivial in the sense that
the Schwinger term in the commutator of the Gauss law generators
is removed by a redefinition of the generators. Indeed, the
Fourier transformed abelian Gauss law generators
\[
\hat{\rm G}_{\pm}(n) \equiv i\frac{2\pi}{\rm L} n \hat{\Pi}_{1,\pm}(n)
+ e_{\pm} \hat{j}_{\pm}(n)
\]
fulfil the algebra
\[
[\hat{\rm G}_{\pm}(n), \hat{\rm G}_{\pm}(m)]_{-} =
\pm e_{\pm}^2 n \delta_{n,-m}.
\]
If we modify the generators as
\[
\hat{\rm G}_{\pm}(n) \rightarrow \hat{\tilde{\rm G}}_{\pm}(n) =
\hat{\rm G}_{\pm}(n) \mp e_{\pm}^2 \frac{\rm L}{4\pi} A_{1,\pm}(n),
\]
then the modified generators commute
\[
[\hat{\tilde{\rm G}}_{\pm}(n),\hat{\tilde{\rm G}}_{\pm}(m)]_{-}=0.
\]
This allows us to define physical states as those which are
annihilated by the modified Gauss law generators,
$\hat{\tilde{\rm G}}_{\pm}(n)|{\rm phys} \rangle =0$
~\cite{sara91}.

The chiral Schwinger model is an exceptional case of models with
anomaly. In contrast with the chiral Schwinger model , the
anomalous behaviour of the chiral $QCD_2$ as well as other 
models with anomaly is non--trivial, i.e.
the Schwinger term in ~\ref{eq: shest} can not be removed. 

To demonstrate this for the chiral $QCD_2$, let us modify the
nonabelian Gauss law generators in the same way as before the
abelian ones for the chiral Schwinger model:
\[
\hat{\rm G}_{\pm}^a(n) \rightarrow \hat{\tilde{\rm G}}_{\pm}^a(n)=
\hat{\rm G}_{\pm}^a(n) + \alpha_{\pm} e_{\pm}^2 A_{1,\pm}^a(n)
\]
where ${\alpha}_{\pm}$ are arbitrary constants.

The commutator algebra for the modified generators is
\[
[\hat{\tilde{\rm G}}_{\pm}^a(n), \hat{\tilde{\rm G}}_{\pm}^b(m)]_{-}=
ie_{\pm} \varepsilon_{abc} \hat{\tilde{\rm G}}_{\pm}^c(n+m)
+i{\alpha}_{\pm} e_{\pm}^3 \varepsilon_{abc} A_{1,\pm}^c(n+m)
\pm e_{\pm}^2 n \delta_{n,-m} \delta^{ab} (1 \pm \frac{4\pi}{\rm L}
\alpha_{\pm} ).
\]
In the right-hand side of this equation, the second term is a new,
additional Schwinger term . This term does not appear in the case
of the chiral Schwinger model where the Gauss law generatores are
scalars. We can not choose ${\alpha}_{\pm}$ in such a way that both
old and new Schwinger terms vanish. For example, if we put , as 
before for the chiral Schwinger model, ${\alpha}_{\pm} =
\mp \frac{\rm L}{4\pi}$, then the old Schwinger term vanishes,
while the new one survives.

For the chiral $QCD_2$, to constrain physical states we act
in another way. Let us note that the Gauss law constraints have
a natural complex structure which relates the positive and negative
Fourier modes:
\[
\hat{\rm G}_{a,\pm}^{\dagger}(n) = \hat{\rm G}_{a,\pm}(-n).
\]
In analogy with the Gupta--Bleuler quantization of ordinary
electrodynamics we require that the physical states are annihilated
only by 'half' of the Gauss law generators ~\cite{fadd86,mick90}.
More precisely, we impose the constraints with positive Fourier index
on the physical ket states
\begin{equation}
\hat{\rm G}_{a,\pm}(n) |{\rm phys}\rangle =0
\hspace{1 cm}  {\rm for} \hspace{1 cm} n \geq 0. \\
\label{eq: sem}
\end{equation}
Then for the constraints with negative Fourier index we have
\[
\langle {\rm phys}| \hat{\rm G}_{\pm}^a(-n) =0
\hspace{1 cm} {\rm for} \hspace{1 cm} n \geq 0,
\]
and therefore all expectation values of the constraints vanish
on the physical states,
\[
\langle {\rm phys}| \hat{\rm G}_{a,\pm}(n) |{\rm phys}\rangle =0
\hspace{1 cm} {\rm for} \hspace{1 cm} {n \in \cal Z}.
\]

Eq.~\ref{eq: cet} implies also that
\[
[\hat{\rm G}_{\pm}^a(n) , \hat{j}_{\pm}^b(m)]_{-} =
ie_{\pm} {\varepsilon}_{abc} \hat{j}_{\pm}^c(n+m) \pm
n e_{\pm} \delta^{ab} \delta_{n,-m} ,
\]
i.e. the fermionic currents no longer have the classical commutator
relations with the Gauss law generators and therefore do not transform
covariantly under gauge transformations.

However, the normal ordering $: {\cdots} :$ is unique only up to
finite terms. There are polynomials in $A_{1,\pm}^a$ which can be
added to the normal ordered fermionic bilinears to make them gauge--
invariant ( the so--called gauge covariant normal ordering
~\cite{lang94} ). In particular, we can define the modified current
operators
\begin{equation}
\hat{\tilde{j}}_{\pm}^a(n) = \hat{j}_{\pm}^a(n) \mp
\frac{\rm L}{2\pi} e_{\pm} A_{1,\pm}^a(n) \\
\label{eq: vosem}
\end{equation}
which obey the desired relations
\[
[\hat{\rm G}_{\pm}^a(n) , \hat{\tilde{j}}_{\pm}^b(m)]_{-} =
i e_{\pm} {\varepsilon}_{abc} \hat{\tilde{j}}_{\pm}^c(n+m),
\]
i.e. have canonical properties under gauge transformations.

Similarly, the Fourier components of the fermionic Hamiltonian
density
\[
\hat{\cal H}_{\rm F}^{\pm}(n) = \hat{\cal H}_{0}^{\pm}(n) \mp
e_{\pm} \sum_{m \in \cal Z} A_{1,\pm}^a(m+n) \hat{j}_{\pm}^a(-m)
\]
do not commute with the Gauss law generators,
$[\hat{\rm G}_{\pm}^a(n) , \hat{\cal H}_{\rm F}^{\pm}(m)]_{-} =
- e_{\pm}^2 n A_{1,\pm}^a(m+n)$ , but the modified ones
\[
\hat{\tilde{\cal H}}_{\rm F}^{\pm}(n) = \hat{\cal H}_{\rm F}^{\pm}(n)
+ \hat{\cal M}^{\pm}(n),
\]
where
\[
\hat{\cal M}^{\pm}(n) \equiv \frac{\rm L}{4\pi} e_{\pm}^2
\sum_{m \in \cal Z} A_{1,\pm}^a(m+n) A_{1,\pm}^a(-m),
\]
are gauge--invariant. We see that the gauge covariant normal ordering
of the fermionic Hamiltonian produces the mass terms
$\hat{\rm M}^{\pm} \equiv \hat{\cal M}^{\pm}(0)$ for the YM fields.

The Fourier components of the YM Hamiltonian density are 
\[
\hat{\cal H}_{\rm YM}^{\pm}(n) = \frac{1}{2\rm L} \sum_{m \in \cal Z}
\hat{\Pi}_{1,\pm}^a(m+n) \hat{\Pi}_{1,\pm}^a(-m).
\]
As known, the operators $\hat{\cal H}_{\rm F}^{\pm}(n)$ and
$\hat{\cal H}_{\rm YM}^{\pm}(n)$ do not have a common, dense invariant
domain of definition in the Hilbert space. It is, however, possible
to define the sum of these operators and therefore the total
Hamiltonian density, if we impose on the vacuum $|{\rm vac};A \rangle$
the condition
\[
(i\hat{\Pi}_{1,\pm}^a(n) + \frac{e_{\pm} \rm L}{\sqrt{2\pi}}
A_{1,\pm}^a(n))|{\rm vac};A \rangle =0 \hspace{5 mm}
{\rm for} \hspace{5 mm} n \in \cal Z,
\]
and order also the YM Hamiltonian density and all other YM field
operators with respect to this vacuum ~\cite{lang94}.

The total gauge--invariant Hamiltonian densities become
\[
\hat{\cal H}^{\pm}(n) = \hat{\cal H}_{0}^{\pm}(n) \mp e_{\pm}
{\vdots} \sum_{m \in \cal Z} A_{1,\pm}^a(m+n) \hat{j}_{\pm}^a(-m)
\]
\begin{equation}
+ \frac{1}{2\rm L} \sum_{m \in \cal Z} (\hat{\Pi}_{1,\pm}^a(m+n)
\hat{\Pi}_{1,\pm}^a(-m) + \frac{e_{\pm}^2{\rm L}^2}{2\pi}
A_{1,\pm}^a(m+n) A_{1,\pm}^a(-m)) {\vdots} , \\
\label{eq: devet}
\end{equation}
where $\vdots \cdots \vdots$ denote the normal ordering for the
YM field operators.

The quantum analogues of the conditions ~\ref{eq: dva} are
\[
\hat{\tilde{\cal H}}_{\rm F}^{\pm}(\frac{\rm L}{2}) |{\rm phys}
\rangle = \hat{\cal H}_{\rm YM}^{\pm}(\frac{\rm L}{2})
|{\rm phys} \rangle =0 ,
\]
or, equivalently,
\begin{eqnarray}
\sum_{n \in \cal Z} (-1)^n \hat{\tilde{\cal H}}_{\rm F}^{\pm}(n)
|{\rm phys} \rangle & = & 0, \nonumber  \\
\sum_{n \in \cal Z} (-1)^n \hat{\cal H}_{\rm YM}^{\pm}(n)
|{\rm phys} \rangle & = & 0.
\label{eq: deset}
\end{eqnarray}

The Sugawara construction ~\cite{godd86} allows to write the
free fermionic Hamiltonian densities $\hat{\cal H}_{0}^{\pm}$
in terms of the Kac--Moody currents $\hat{j}_{\pm}^a$,
\[
\hat{\cal H}_{0}^{\pm}(n) = \frac{\pi}{\rm L}
\sum_{n \in \cal Z} \stackrel{\times}{\times}
\hat{j}_{\pm}^a(n+m) \hat{j}_{\pm}^a(-m)
\stackrel{\times}{\times}
\]
with normal ordering $\stackrel{\times}{\times} \hat{j}_{\pm}^a(k)
\hat{j}_{\pm}^b(q) \stackrel{\times}{\times} \equiv
\hat{j}_{\pm}^b(q) \hat{j}_{\pm}^a(k)$ for $k \stackrel{<}{>} q$
and $\hat{j}_{\pm}^a(k) \hat{j}_{\pm}^b(q)$ otherwise. Note that
$\hat{j}_{+}(k)|{\rm vac};{\rm F} \rangle =
\hat{j}_{-}(-k)|{\rm vac};{\rm F} \rangle =0$ for $k>0$.
Combining this with ~\ref{eq: vosem}, we finally get the Hamiltonian
of the model in the following form
\[
\hat{\rm H} = \frac{\pi}{\rm L} \vdots \sum_{n \in \cal Z}
\{ \stackrel{\times}{\times} ( \hat{\tilde{j}}_{+}^a(n)
\hat{\tilde{j}}_{+}^a(-n) + \hat{\tilde{j}}_{-}^a(n)
\hat{\tilde{j}}_{-}^a(-n) ) \stackrel{\times}{\times}
\]
\begin{equation}
+ \frac{1}{2\pi} ( \hat{\Pi}_{1,+}^a(n) \hat{\Pi}_{1,+}^a(-n)
+ \hat{\Pi}_{1,-}^a(n) \hat{\Pi}_{1,-}^a(-n) ) \} \vdots .
\label{eq: odinodin}
\end{equation}

$4.$ The quantum momentum and boost generators are
\[
\hat{\rm P}_{\pm} = \pm \hat{\cal H}_{\rm F}^{\pm}(0) +
\sum_{n>0} \hat{\rm G}_{\pm}^a(-n) {A}_{1,\pm}^a(n) +
\sum_{n \leq 0} {A}_{1,\pm}^a(n) \hat{\rm G}_{\pm}^a(-n),
\]
\[
\hat{\rm K}_{\pm} = -i \frac{\rm L}{2\pi} \sum_{\stackrel{n \in \cal 
Z}
{n \neq 0}} \frac{1}{n} (-1)^n \hat{\cal H}^{\pm}(n).
\]
Commuting $\hat{\rm P}_{\pm}$ and $\hat{\rm G}_{\pm}^a(n)$, we get
\[
[\hat{\rm G}_{\pm}^a(n), \hat{\rm P}_{\pm}]_{-} =
\frac{2\pi}{\rm L} n \hat{\rm G}_{\pm}^a(n),
\]
i.e. the quantum momentums are gauge invariant only in the sector
of the physical states ~\ref{eq: sem}. Moreover,
\[
\langle {\rm phys}| \hat{\rm P}_{\pm}|{\rm phys} \rangle =
\pm \langle {\rm phys}| \hat{\cal H}_{\rm F}^{\pm}(0)|
{\rm phys} \rangle .
\]

Now, we construct the algebra of the quantum Hamiltonian $\hat{\rm 
H}$,
the momentum $\hat{\rm P} =\hat{\rm P}_{+} + \hat{\rm P}_{-}$ and
the boost generator $\hat{\rm K} = \hat{\rm K}_{+} +
\hat{\rm K}_{-}$. With ~\ref{eq: deset}, it is straightforward to
check that on the physical states
\[
[\hat{\rm H} , \hat{\rm P}]_{-}  =  0,
\]
\[
[\hat{\rm P} , \hat{\rm K}]_{-} =  -i \hat{\rm H},
\]
and
\[
[\hat{\rm H} , \hat{\rm K}]_{-} = -i \hat{\rm P} +
i (\hat{\rm M}^{-} - \hat{\rm M}^{+} ).
\]
We see that these commutation relations differ from those of the 
Poincare algebra. The difference is in the mass terms in the 
last commutator. Only in the case of the standard $QCD_2$, 
$\hat{\rm M}^{+}=\hat{\rm M}^{-}$ and we get the Poincare algebra. 

The mass terms $\hat{\rm M}^{\pm}$ can not be removed from the
algebra by a redefinition of the Poincare generators, if the
generators are required to be gauge invariant. We have added
these mass terms to the fermionic and total Hamiltonians just
to make them gauge invariant.

Thus, for the chiral $QCD_2$ the Poincare algebra fails to close
in the physical sector where the states satisfy the constraints
~\ref{eq: sem} and the Poincare generators are gauge invariant.
We have constructed the commutation relations of the new algebra
explicitly in a compact form.

The failure  of the Poincare algebra to close on the physical states
implies that the model is not relativistically invariant. The
physical Hamiltonian and momentum commute,so translational invariance
is preserved. This situation is similar to that in the chiral
Schwinger model (see, for example, \cite{hall86},\cite{sara91},
\cite{niemi86}). The analysis performed in these references shows
that when we construct a quantum theory consistent with the anomaly
and use the Gauss law to constrain physical states, then relativistic
invariance is lost. In other words, the Poincare algebra fails
to close on the physical states for the chiral Schwinger model, too. 

The origin of the breakdown of relativistic invariance is 
the same in both models and lies in the anomaly. Therefore, for
the chiral $QCD_2$ as well as for the chiral Schwinger model
the anomaly or ,equivalently, the fact that the local gauge symmetry
is realized projectively disturbs relativistic invariance. We
believe that this is a fundamental feature characteristic for
anomalous models. However, the question of whether relativistic
invariance is broken for other models with the projective
realization of a local gauge symmetry, especially in higher 
dimensions,
remains open.

\end{document}